\documentclass[twocolumn,showpacs,superscriptaddress]{revtex4}

\usepackage{amsmath,amssymb}
\usepackage{graphicx}
\usepackage{dcolumn}
\usepackage{bm}
\usepackage{xcolor}
\begin{document}

\title{All-optical reversible logic gate via adiabatic population transfer}

\author{G. Grigoryan}
\email{gaygrig@gmail.com}\affiliation{Institute for Physical
Research, Armenian National Academy of Sciences, Ashtarak-2, 0203,
Armenia} \affiliation{Russian-Armenian (Slavonic) University}
\author{V. Chaltykyan}
\affiliation{Institute for Physical Research, Armenian National
Academy of Sciences, Ashtarak-2, 0203, Armenia}
\author{E. Gazazyan}
\affiliation{Institute for Physical
Research, Armenian National Academy of Sciences, Ashtarak-2, 0203,
Armenia}
\author{O. Tikhova}
\affiliation{Institute for Physical
Research, Armenian National Academy of Sciences, Ashtarak-2, 0203,
Armenia}
\author{T. Halfmann}
\affiliation{Institute of Applied Physics, Technical University of Darmstadt, Hochschul{stra\ss e~6}, 64289 Darmstadt, Germany}
\date{\today}

\begin{abstract}
The Toffoli gate is  an essential logic element, which permits implementation of a reversible processor. It is of relevance both for classical as well as quantum logics. We propose and theoretically study all-optical implementations of three-bit and four-bit Toffoli gates by application of adiabatic population transfer techniques. For a three-bit Toffoli gate we use variants of stimulated Raman adiabatic passage (STIRAP) processes in a $\Lambda$-type level scheme, driven by two laser pulses at sufficiently large detunings. For the implementation of a four-bit Toffoli gate, we apply reversible adiabatic population transfer in five-level quantum systems, interacting with three laser pulses. We demonstrate correct all-optical implementation of the truth table of three-bit and four-bit Toffoli gates. Moreover, we derive conditions for adiabatic evolution of the population dynamics and robust operation of the gates.
\end{abstract}

 \maketitle

\section{\protect\normalsize INTRODUCTION}
Reversible and irreversible computations exhibit very different features with regard to energy consumption. As it was already shown decades ago by Landauer, irreversible computations unavoidably lead to energy losses \cite{1 Landauer}. Thus, in contrast to reversible processors they continuously consume energy \cite{2 Bennett,3 Preskil,4 Nielsen}. In 1980, Toffoli proposed a reversible processor \cite{5 Toffoli} based on a reversible logic gate (i.e. the Toffoli gate).  Any reversible processor can be realized by circuits of Toffoli gates only. A general $\it{n}$-bit Toffoli gate has $\it{n}$ input bits and yields an $\it{n}$-bit output. The first ($\it{n}$-1) input bits are control bits, which are  not affected by the action of the Toffoli gate. The last input bit is a target bit, which is flipped, if (and only if) all control bits are set to 1.

Also other universal reversible logic gates were proposed later \cite{3 Preskil,4 Nielsen}. We note, that all quantum logic gates are reversible. As a consequence, a transfer of quantum logic schemes to and from classical logics should be based on reversible units, e.g. Toffoli gates. This makes such gates important both for classical as well as quantum information science \cite{6 Peres, 7 Pachos,8 Duan,9 Chang-Yong Chen,10 Ralph,11 Ionicioiu,12 Fiurasek,13 Lanyon,14 Judson,15 Monz,16 Ospelkaus}.

In this paper we propose and theoretically investigate an all-optical three-bit and four-bit Toffoli gate, based upon coherent-adiabatic interactions between quantum systems (e.g. atoms) and resonant light fields. In particular we apply cyclic stimulated Raman adiabatic passage (STIRAP), driven by two delayed light pulses \cite{17 Vitanov,18 Bergmann,19 Kral}.

The STIRAP scheme \cite{17 Vitanov} uses three non-degenerate bare quantum states in a $\Lambda$-type configuration, i.e. an initial state $|1\rangle$, an intermediate state $|2\rangle$, and a final state $|3\rangle$ (see Fig. 1). A pump pulse with Rabi frequency $\Omega_1$ drives the transition between states $|1\rangle$ and $|2\rangle$. A Stokes pulse with Rabi frequency $\Omega_2$ drives the transition between states $|2\rangle$ and $|3\rangle$. When the laser pulses are applied in counter-intuitive order (i.e. the Stokes pulse precedes the pump pulse) and both Rabi frequencies are sufficiently large, the atomic population is $\it{completely}$ driven via a "dark" dressed state from the initial state $|1\rangle$ to the final state $|3\rangle$. Thus, the system is flipped efficiently from state $|1\rangle$ to state $|3\rangle$. We note, that the "dark" state involves only contributions from states $|1\rangle$ and $|3\rangle$, but not from state $|2\rangle$. As an essential feature of this standard "dark-state" STIRAP with a counter-intuitive pulse sequence, there is never any transient population in the intermediate state $|2\rangle$ during the adiabatic passage process. This is an important issue, if the intermediate state $|2\rangle$ has a short population lifetime compared to the laser pulse durations (which is a rather typical situation).

When state $|2\rangle$ has a long lifetime, it is also possible to transfer population by an intuitive pulse sequence (i.e. pump preceding Stokes pulse) via a "bright" dressed state. The latter involves contributions from all three bare states $|1\rangle$, $|2\rangle$, and $|3\rangle$. This variant of STIRAP with an intuitive pulse sequence was termed $\it{b}$-STIRAP to emphasize the adiabatic passage via a bright state \cite{20 Klein07,21Klein08}. Such a transfer process in an individual atomic system was predicted and studied theoretically by Rangelov et al. \cite{22 A. Rangelov}. For an ensemble of atoms we theoretically analyzed the effect in previous work \cite{23 Grigoryan09}. Experimentally, $\it{b}$-STIRAP was demonstrated for population transfer in a doped solid \cite{20 Klein07,21Klein08}.

We note, that the definition of an intuitive or counter-intuitive pulse sequence depends upon the state of the atom at the beginning of the interaction. If the atom is initially in state $|1\rangle$, the pulse sequence "Stokes preceding pump" is counter-intuitive. In this case, the sequence transfers the system to state $|3\rangle$ via standard (i.e. "dark-state") STIRAP. If the atom is initially in state $|3\rangle$, the roles of the laser pulses are exchanged. Thus, the pulse sequence "Stokes preceding pump" becomes intuitive. In this case, the sequence transfers the system to state $|1\rangle$ via $\it{b}$-STIRAP.

Let us call the sequence "Stokes preceding pump" a "SP pulse pair". We consider now application of two subsequent SP pairs to the three-level quantum system. If the atom is initially in state $|1\rangle$, the first SP pair flips the system to $|3\rangle$ by STIRAP. The second SP pair flips the system back to state $|1\rangle$ by $\it{b}$-STIRAP. A similar transfer back and forth occurs, when the atom is initially in state $|3\rangle$. Thus, population transfer by two SP pairs in a $\Lambda$-type three-level system is fully reversible. The same pulse sequence (i.e. an SP pair) serves to drive atomic population from the initial state to the final state or back again. As we will show below, this feature enables implementation of a three-bit Toffoli gate. The possibility to exploit STIRAP for classical logics was initially proposed theoretically by Remacle et al. \cite{24 Remacle}. The concepts were experimentally applied to implement an all-optical adder in a doped solid, driven by STIRAP and $\it{b}$-STIRAP \cite{25 Beil}.

In the following we will provide a detailed and general study of adiabatic passage processes for reversible universal logic operators. Although any reversible logic processor could be built with three-bit units only, the processor would be rather complex and large. Thus, it is very advisable to apply four-bit Toffoli units instead \cite{26Golubitsky}. As we will discuss, below, we can implement such a four-bit gate by adiabatic passage processes in a five-level system, driven by three laser pulses representing the input bits (see Fig. 2). We note, that such interaction schemes are possible in the same doped solids, as applied by Beil et al. for a STIRAP-driven all-optical adder \cite{25 Beil}. Coherent population transfer in multi-level systems was already studied  before \cite{27 Ottaviani,28 Mollmer,29 Kuznetsova,30 Amniat-Talab,31 T.Noel,32 Yatsenko}. As an example, in M-type five-level schemes (see Fig. 2(a)) efficient population transfer is possible via chains of STIRAP processes \cite{17 Vitanov}.We propose now an alternative technique for efficient population transfer in five-level schemes, based upon combination of STIRAP and $\it{b}$-STIRAP. The scheme is completely reversible and enables realization of a Toffoli gate.

The paper is organized as follows : In section II we discuss a three-bit Toffoli gate in a $\Lambda$-type system. In section III we study cyclic population transfer in five-level systems. Based upon these results, in section IV we propose the implementation of a four-bit Toffoli gate in an adiabatically-driven  medium of five-level atoms. We conclude with a final discussion in section V.

\begin{figure}
 \includegraphics[width=6cm]{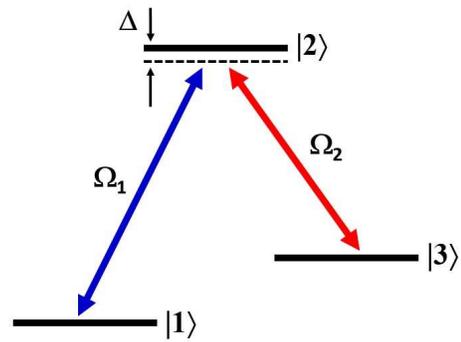}\\
  \caption{$\Lambda$-type level scheme}\label{1u1}
\end{figure}
\begin{figure}
  \includegraphics[width=8.5cm]{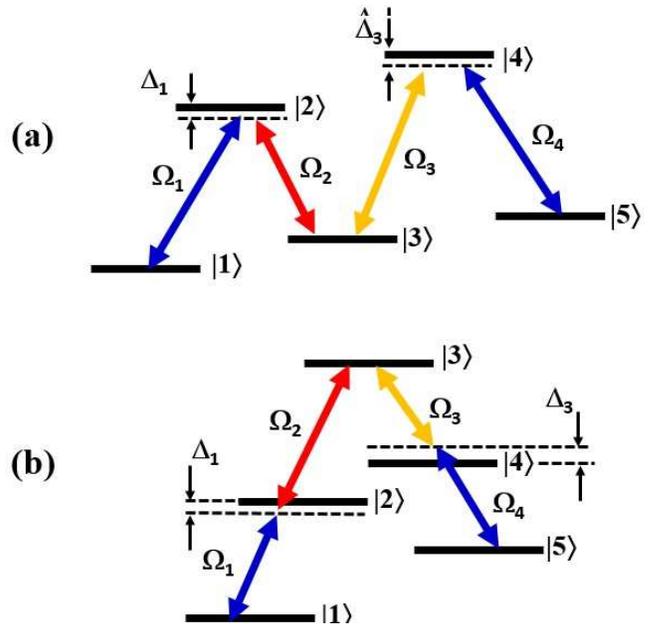}\\
  \caption{Five-level coupling schemes : (a) M-type scheme. (b) Extended $\Lambda$-scheme.}\label{1u1}
\end{figure}

\section{Realization of a three-bit Toffoli gate in a three-level $\Lambda$-system}
The well-known dynamics of a coherently-driven three-level $\Lambda$-type quantum system (see Fig.1) yields three eigenvectors, i.e. one dark and two bright dressed states \cite{18 Bergmann}:
\begin{eqnarray}\label{eq1.1}
|d\rangle = cos\theta e^{-i\varphi_1}|1\rangle-sin\theta e^{-i\varphi_2}|3\rangle \nonumber \\
|b1\rangle = sin\theta cos\Phi e^{-i\varphi_1}|1\rangle-sin\Phi |2\rangle+cos\theta cos\Phi e^{-i\varphi_2}|3\rangle  \nonumber \\
|b2\rangle = sin\theta sin\Phi e^{-i\varphi_1}|1\rangle+cos\Phi |2\rangle+cos\theta cos\Phi e^{-i\varphi_2}|3\rangle
\end{eqnarray}
The mixing angles are defined as
\begin{eqnarray}\label{eq1.2}
tan \theta=\Omega_1/\Omega_2 \nonumber \\
tan 2\Phi=2\Omega/\Delta\nonumber\\
\Omega=\sqrt{\Omega^2_1+\Omega^2_2} \nonumber \\
\end{eqnarray}
where $\Delta$ is the one-photon detuning, $\Omega_{1,2}$ are the Rabi frequencies of pump and Stokes laser,
and $\varphi_{1,2}$ are the phases of the laser pulses.
In the above equations we assumed the lasers tuned to two-photon resonance between states $|1\rangle$ and $|3\rangle$.
We note, that the Rabi frequencies (and hence the mixing angles) vary with time $t$ and space coordinate $x$,
as they are deduced from temporal laser intensity profiles, which also vary when propagating through an optically dense medium.

If the atom is initially in state $|1\rangle$ and we apply a counter-intuitive pulse sequence (i.e. a SP pair), the mixing angle yields $\theta=0$ and the system is prepared in the dark state $|d\rangle$. Provided we maintain adiabaticity (i.e. the system remains during the interaction in the same dressed state), at the end of the interaction the mixing angle becomes $\theta=\pi/2$ and the dark state maps onto the final state $|3\rangle$. This is the essence of STIRAP. If the atom is initially in state $|3\rangle$ and we apply an intuitive pulse sequence (i.e. also a SP pair, as the laser pulses change their role now), the system aligns  parallel with a bright state $|b1\rangle$. Provided we maintain adiabaticity, at the end of the interaction the bright state maps onto state $|1\rangle$. This is the essence of $\it{b}$-STIRAP. In both variants of STIRAP, adiabatic evolution requires the following adiabaticity condition to be fulfilled \cite{17 Vitanov,18 Bergmann,19 Kral}:
\begin{eqnarray}\label{eq1.3}
|\Omega^2T/\Delta|\gg1
\end{eqnarray}
with the duration $T$ of the interaction, e.g. defined by the laser pulse duration.
Essentially, adiabaticity demands sufficiently strong interaction (defined by the Rabi frequency) compared to the laser bandwidth (defined by the Fourier bandwidth ~$1/T$) and the detuning. To realize a Toffoli gate (see below) it is also required to maintain the adiabaticity condition for a two-level medium
\begin{eqnarray}\label{eq1.31}
\Delta T\gg1 \nonumber
\end{eqnarray}
i.e., the one-photon detuning should exceed well the spectral width of lasers.
We note, that pulse propagation effects (e.g. in optically thick media) require modifications of the adiabaticity criteria. Detailed analysis showed, that propagation effects may be neglected, if the medium satisfies the conditions \cite{22 A. Rangelov,23 Grigoryan09}.
\begin{eqnarray}\label{eq1.4}
qL/\Omega^2 T\ll1 \nonumber \\
qL/\Delta^2 T\ll1
\end{eqnarray}
with the length $L$ of the medium, the coupling parameter $q$ defined as q=max$[q_1,q_2]$, $q_{1,2}=2\pi\omega_{1,2}d^2_{1,2} N/\hbar c $,
involving the transition dipole moments $d_{1,2}$, the transition frequencies $\omega_{1,2}$, and the atomic number density $N$ in the medium.
Thus, the optical length of the medium should remain sufficiently small.
At short optical lengths, adiabaticity of the interaction is preserved and efficient population transfer is possible.

To implement a three-bit Toffoli gate in a $\Lambda$-type interaction scheme, we apply a pump and a Stokes pulse in counter-intuitive sequence (i.e. a SP pair). We assume pulse durations much shorter compared to relaxation times in the system.
The laser frequencies are tuned to two-photon resonance between states $|1\rangle$ and $|3\rangle$, but both lasers are detuned by $\Delta$ from the transition to the intermediate state $|2\rangle$. The two laser pulses define the first and second input of the gate (see truth table).

If the pump pulse is switched on, it defines the logic value 1 of the first input bit. In the same way, presence or absence of the Stokes pulse defines the second input bit. The third input of the gate is defined by the state of the atom before the interaction. If the atom is in state $|1\rangle$, the third input of the gate is equal to the logic value 0. If the atom is in state $|3\rangle$, the third input of the gate is equal to the logic value 1. We consider now the following possible cases : (A) If there are no pulses applied, the atom remains in the initial state (see the first two lines in the truth table). (B) If the pump pulse is switched off and only the Stokes pulse is applied, an atom prepared in state $|1\rangle$ will remain in this state (see third line in the truth table). An atom prepared in state $|3\rangle$ will be adiabatically driven by the off-resonant Stokes pulse to the intermediate state $|2\rangle$ and back again. This effect is termed "coherent population return" (CPR) (see \cite{41 Chakrabarti,40 Peralta} and refs. therein).
Thus, also in this case the final state of the atom is the same as the initial state (see fourth line in the truth table).(C) If the Stokes pulse is switched off and only the pump pulse is applied, the dynamics are similar to the previous case. Also here, the state of the atom does not change (see fifth and sixth line in the truth table). (D) If both pulses are applied, an atom in state $|1\rangle$ is driven by STIRAP to state $|3\rangle$(see seventh line in the truth table). An atom in state $|3\rangle$ is driven by $\it{b}$-STIRAP to state $|1\rangle$ (see eighth line in the truth table). Thus, the truth table of the population dynamics mirrors the logics of a three-bit Toffoli gate.
We note, that for experimental implementation there are several ways to determine the output state of the gate, e.g., detection of fluorescence via an additional probe laser.
\begin{table}
\centering
\begin{tabular}{|c|c|c||c|}
   \hline
 \multicolumn{3}{|c||}{input} &  output \\ \hline \hline
  pump & stokes & initial state & final state \\ \hline \hline
  0 & 0 & 0 & 0 \\ \hline
  0 & 0 & 1 & 1 \\ \hline
  0 & 1 & 0 & 0 \\ \hline
  0 & 1 & 1 & 1 \\ \hline
  1 & 0 & 0 & 0 \\ \hline
  1 & 0 & 1 & 1 \\ \hline
  1 & 1 & 0 & 1 \\ \hline
  1 & 1 & 1 & 0 \\ \hline
\end{tabular}
\caption{Truth table of a three-bit Toffoli gate, with input bits defined by pump pulse, Stokes pulse, and the initial atomic state. The output state is defined by the final atomic state.} \label{t1}
\end{table}

\section{Population dynamics in five-level schemes}
To proceed towards a more complex system and a gate with more input bits, we consider now a five-level system interacting with four pulses (see Fig. 2). In particular, we consider an M-system (see Fig. 2 (a)) and an extended $\Lambda$-system involving two-photon transitions on the two branches (see Fig. 2(b)).
M-systems are typical, e.g. for atoms with hyperfine or Zeeman splittings. The extended $\Lambda$-system occurs for transitions to highly excited states in atoms or molecules.
The laser frequencies are tuned near resonance with one of adjacent atomic transitions. We assume pulse durations much shorter compared to relaxation times in the system.

The Hamiltonian yields
\begin{equation}\label{eq1}
    H=\sum_i\sigma_{i,i}\delta_{i-1}-\biggr(\sum_i\sigma_{i,i+1}\Omega_i+h.c.\biggl)
\end{equation}
with the projection matrices $\sigma_{ii}$, the Rabi frequencies $\Omega_i$ at transitions ($i\rightarrow i$+1), and $\delta_{i-1}$ representing ($i$-1)-photon detunings (with $\delta_0=0$). The Rabi frequencies are assumed to be real and positive. Phases, which can  vary during propagation through the medium, are included in the single-photon detunings ($\Delta_i=\omega_{i+1,i}-\omega_i+\dot{\varphi}_i$, if $\omega_{i+1,i}>0$ and $\Delta_i=\omega_{i,i+1}-\omega_i+\dot{\varphi}_l$ if $\omega_{i+1,i}<0$). Definition of multi-photon detunings depends on the specific scheme of interaction. For an M-system (see Fig. 2(a)) the multi-photon detunings are $\delta_2=\Delta_1-\Delta_2,\ \delta_3=\Delta_3+\Delta_1-\Delta_2, \ \delta_4=\Delta_4-\Delta_3+\Delta_2-\Delta_1$. For an extended $\Lambda$-system (see Fig. 2(b)), the multi-photon detunings are $\ \delta_2=\Delta_1+\Delta_2,\ \delta_3=-\Delta_3+\Delta_1+\Delta_2, \ \delta_4=-\Delta_4-\Delta_3+\Delta_2+\Delta_1$.

Similar to the $\Lambda$ system, we will assume now exact two photon resonances, i.e. $\delta_2=\delta_4=0$ and $\delta_1=\delta_3=\Delta$. For an M-system this condition means equal single-photon detunings, while for the extended $\Lambda$-scheme the single-photon detunings have equal absolute value, but differ in sign (see Fig.2 (b)). In this case, one of five eigenvalues of the Hamiltonian is $\lambda=0$. We can also easily calculate the other four eigenvalues (see Appendix A).

We consider now the specific case, when the pulses with $\Omega_1$ and $\Omega_4$ coincide (i.e. exhibit the same temporal profile and equal frequencies), while the pulses with $\Omega_2$ and $\Omega_3$ are much shorter and turned on in counterintuitive sequence (see Fig.3).
In this case, the eigenvalues of the Hamiltonian (1) are
\begin{eqnarray}\label{eq2}
\Lambda_0=0,\nonumber \\
\Lambda_{1,3}=\frac{1}{2}\biggr(\Delta\mp\sqrt{\Delta^2+4\Omega^2_1}\biggl),\nonumber\\
\Lambda_{2,4}=\frac{1}{2}\biggr(\Delta\mp\sqrt{\Delta^2+4(\Omega^2_1+\Omega^2_2+\Omega^2_3)}\biggl)
\end{eqnarray}
We note, that when the fields are turned off, we get $\Lambda_{1,2}\rightarrow 0$ and $\Lambda_{3,4}\rightarrow\Delta.$  The eigenvalues $\Lambda_{1,3}$ depend upon the field $\Omega_1$ only and coincide with the eigenvalues of a two-level system, driven by field $\Omega_1$. Similarly, the eigenvalues $\Lambda_{2,4}$ are equal to the eigenvalues of a two-level system, driven by a field $(\Omega^2_1+\Omega^2_2+\Omega^2_3)^{1/2}$.

When the one-photon detuning is larger (or at least of the same order of magnitude) as the corresponding Rabi frequencies, adiabatic evolution requires the following conditions to be met (see Appendix A for details):
\begin{eqnarray}\label{eq3}
\Delta T\gg1,\ \ \frac{(\Omega^2_2+\Omega^2_3)T}{\Delta}\gg1,\ \ \frac{\Omega^2_1T}{\Delta}\gg1
\end{eqnarray}
with the duration $T$ of the shortest pulse. The first condition mirrors the adiabaticity condition for a two-level system. The second condition corresponds to the adiabaticity condition for a three-level system. The third condition is only relevant in the time interval, when all pulses overlap (i.e., when $\Omega^2_2+\Omega^2_3\neq0$).

To write the eigenvectors corresponding to the eigenvalues  $\Lambda_1$ and $\Lambda_2$ we introduce the following notations:
\begin{eqnarray}\label{eq4}
\Omega^2=\Omega^2_2+\Omega^2_3,\ \ \tan\theta=\frac{\Omega_2}{\Omega_3}, \nonumber \\
\tan{\Phi_1}=-\frac{\Lambda_1}{\Omega_1}, \tan{\Phi_2}=-\frac{\Lambda_2}{\Omega_1},\ \tan{\Phi}=-\frac{\Omega}{\Omega_1}\cos{\Phi_2}
\end{eqnarray}

The eigenvector corresponding to the eigenvalue $\Lambda_1$ is
\begin{equation}\label{eq5}
|\Lambda_1\rangle=|\psi_1\rangle\cos{\theta}-|\psi_2\rangle\sin{\theta}
\end{equation}
where $|\psi_1\rangle$ and $|\psi_2\rangle$ are superposition states of two-level systems $1\rightarrow2$ and $5\rightarrow4$:
\begin{eqnarray}\label{eq6}
|\psi_1\rangle=\cos{\phi_1}|1\rangle-\sin{\phi_1}|2\rangle \nonumber\\
|\psi_2\rangle=\cos{\phi_1}|5\rangle-\sin{\phi_1}|4\rangle
\end{eqnarray}
The eigenvector corresponding to $\Lambda_1$ does not involve state $|3\rangle$ and is equal to the dark state of a three-level $\Lambda$-system, if we replace the lower states by superposition states $|\psi_1\rangle$ and $|\psi_2\rangle$. Thus, we can use the dark state to transfer the system from state $|1\rangle$ state $|5\rangle$ by a STIRAP-like process, driven
by the pulse sequence introduced above (see Fig.3).
In contrast to a simple three-level $\Lambda$-system, during the interaction some transient population shows up in the intermediate levels $|2\rangle$ and $|4\rangle$ of the five-level system. However, these transient populations become smaller with larger single-photon detuning.
In this case, the intermediate states $|2\rangle$ and $|4\rangle$ mediate a coupling between states $|1\rangle$, $|3\rangle$, and $|5\rangle$, but are only weakly populated during the process.

\begin{figure}[h]
 input(1110)$\longrightarrow$ output(1111)
 \includegraphics[width=8cm]{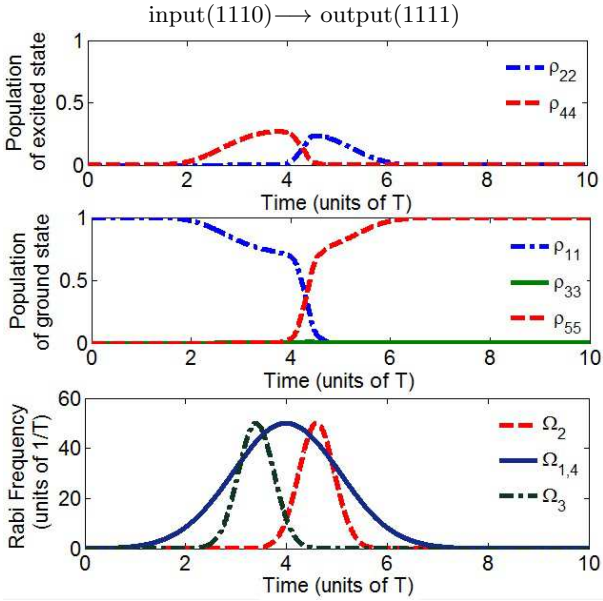}\\
  \caption{Adiabatic population transfer from initial state $|1\rangle$ to final state $|5\rangle$ and pulse sequence.
   $\rho_{jj}$ are the populations of corresponding levels. All pulse shapes are Gaussian. The single-photon detuning is $\Delta$=50/T.}\label{p4}
\end{figure}

\begin{figure}[h]
  input(1111)$\longrightarrow$ output(1110)
  \includegraphics[width=8cm]{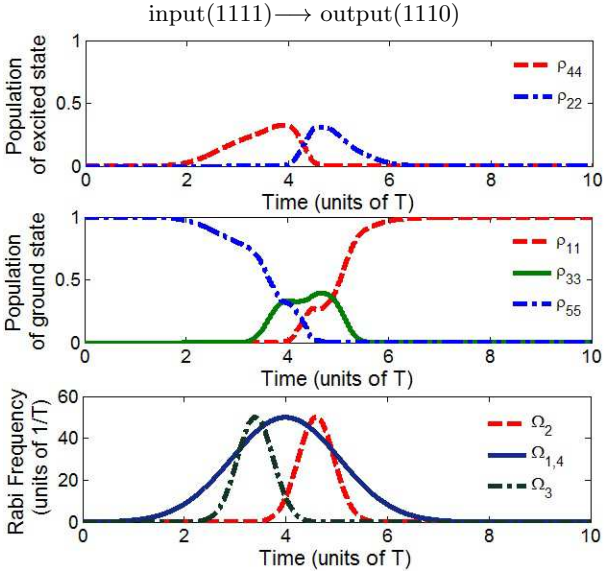}\\
  \caption{Adiabatic population transfer from final state $|5\rangle$ to initial state $|1\rangle$ and pulse sequence. Parameters are the same as in Fig.3.}\label{p5}
\end{figure}

\begin{figure}[h]
  input(1000)$\longrightarrow$ output(1000)
  \includegraphics[width=8cm]{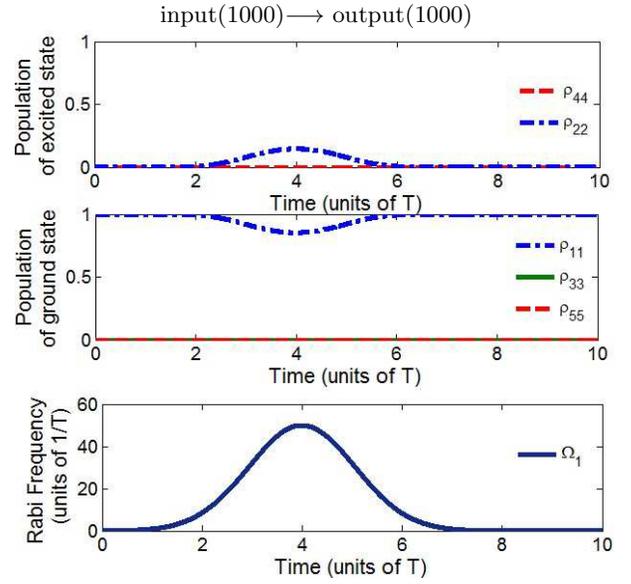}\\
  \caption{Adiabatic evolution (CPR) of populations in a two-level system. Parameters as in previous figures.}\label{p7}
\end{figure}

\begin{figure}[h]
  input(1010)$\longrightarrow$ output(1010)
  \includegraphics[width=8cm]{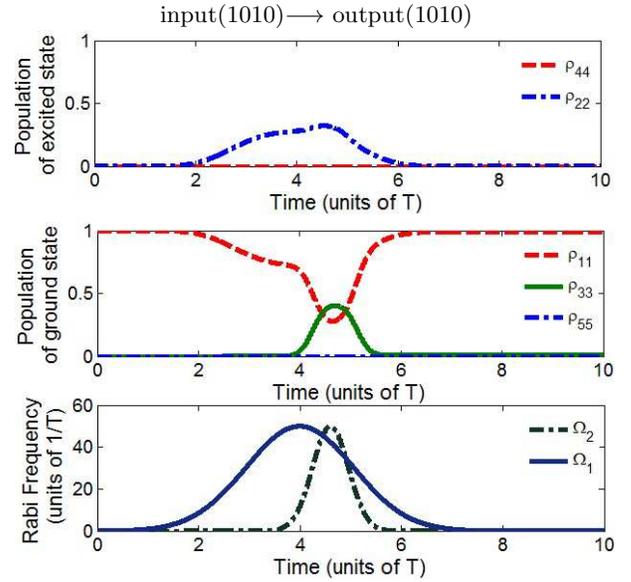}\\
  \caption{Adiabatic evolution of populations in a three-level system. Parameters as in previous figures. }\label{p8}
\end{figure}

%
%

Similarly, the eigenvector corresponding to the eigenvalue $\Lambda_2$ yields
\begin{equation}\label{eq7}
|\Lambda_2\rangle=|\psi'_1\rangle \cos{\Phi}\sin{\theta}-\sin{\phi}|3\rangle +|\psi'_2\rangle \cos{\Phi}\cos{\theta},
\end{equation}
where
\begin{eqnarray}\label{eq8}
|\psi'_1\rangle=\cos{\Phi_2}|1\rangle-\sin{\Phi_2}|2\rangle \nonumber \\
|\psi'_2\rangle=\cos{\Phi_2}|5\rangle-\sin{\Phi_2}|4\rangle
\end{eqnarray}
The eigenvector $|\Lambda_2\rangle$  is equal to that of the bright state of a three-level $\Lambda$-system [24], if we replace the lower states by superposition states $|\psi'_1\rangle$ and $|\psi'_2\rangle$.
Thus, we can use the bright states for adiabatic transfer from state $|5\rangle$  to state $|1\rangle$ by a $\it{b}$-STIRAP-like process, driven by the pulse sequence as introduced above (see Fig. 4).

In order to ensure adiabaticity in the five-level dynamics, the following condition must be added to the adiabaticity conditions (4), as discussed above :
\begin{equation}\label{eq14}
\frac{q_1L}{\Delta}\frac{\Delta}{\Omega^2_1 T}\ll 1
\end{equation}.
The latter is obtained from Maxwell's equations (see Appendix B). It follows from (4) and (13), that the strength of pulse propagation effects upon adiabaticity are determined by the factor $q_i L/\Delta$, combined with the adiabaticity conditions for a single atom. Thus, $q_i L/\Delta$ should not exceed unity. If we write this parameter in terms of the linear absorption coefficient $\alpha_0$, we obtain a restriction for the optical length :
\begin{equation}\label{eq15}
\frac{qL}{\Delta}=\alpha_0 L \frac{\Gamma}{\Delta}\sim 1
\end{equation}
with $\Gamma$ the maximal decay rate  of the involved transitions.
Thus, for sufficiently large single-photon detuning, adiabaticity is maintained even when the medium is several times longer than the linear absorption length in the medium.

In summary, the two eigenstates $|\Lambda_1\rangle$ and $|\Lambda_2\rangle$ render the five-level system, driven by a considered pulse sequence, fully reversible. Thus, we can transfer atomic population from state $|1\rangle$ to state $|5\rangle$ by a STIRAP-like process and from state $|5\rangle$  to state $|1\rangle$ by a $\it{b}$-STIRAP-like process.


\section{Implementation of a four-bit Toffoli gate in a five-level system}

The light fields $\Omega_1$,  $\Omega_2$, and  $\Omega_3$ play the role of the three control bits in the four-bit Toffoli gate. We note, that in our coupling scheme the field $\Omega_1$ is identical to $\Omega_4$ (see previous section). If a field is switched on, the corresponding control input bit is defined with logic value 1. If the field is switched off, the corresponding control input bit is defined with the logic value 0. The fourth input bit of the gate is the target bit. The logic value of the target bit is defined by the states $|1\rangle$  and $|5\rangle$ of the atom. If the atom is in state $|1\rangle$, the target bit is defined as 0. If the atom is in state $|5\rangle$, the target bit is 1. In the following we will denote the input and output state of the gate by the sequence of control bits $c_i $ and target bit $t$, i.e. ($c_1$,$c_2$,$c_3$,$\it{t}$). As discussed in the previous section, simultaneous action of the light fields drives adiabatic population transfer between states $|1\rangle$ and $|5\rangle$. This corresponds to a gate operation from the input state (1110) to the target state (1111), or from the input state (1111) to the output state (1110).

Beyond this feature (as already discussed in the previous section), correct operation of the gate also requires that the state of the atom does not change, if at least one of the driving fields is switched off. Obviously this holds true in the trivial case, when all fields are switched off. Also the case, when only field $\Omega_1$ (which is identical to $\Omega_4$) is rather trivial : Neither state $|1\rangle$ nor state $|5\rangle$ experience any coupling to other states and the atom remains in the initial state. If fields $\Omega_2$ and $\Omega_3$ are both switched off simultaneously, the five-level system reduces to a two-level scheme, adiabatically interacting with pulse  $\Omega_1$. The atom experiences CPR. After some transient excitation the atoms returns to the initial state (see Fig.5).
		
The case where only one of pulses $\Omega_2$ and $\Omega_3$ is switched off, requires some more detailed consideration. If the atom resides in state $|1\rangle$ (i.e. the target bit is 0) and the field  $\Omega_2$ is off, the system reduces to an adiabatically-driven two-level atom. Due to the action of field $\Omega_1$, the atom experiences CPR and can be found in state $|1\rangle$ after the interaction. A similar situation occurs when the atom is in state $|5\rangle$ (i.e. the target bit is 1) and field  $\Omega_3$ is off. If the atom is initially in state $|1\rangle$  and only pulse $\Omega_3$ is switched off, the system reduces to a standard three-level scheme of states $|1\rangle$, $|2\rangle$, and $|3\rangle$, adiabatically driven by fields $\Omega_1$ and $\Omega_2$. As the $\Omega_2$ pulse is turned off prior to $\Omega_1$, the system returns after interaction into the initial state (see Fig.6). We get an equivalent situation, if the atom is initially in state $|5\rangle$  and only pulse $\Omega_2$ is switched off. Also in this case, the atom remains in the initial state. Thus, in summary, the adiabatic population dynamics in the five-level scheme perfectly corresponds to a four-bit Toffoli gate.

\section{\protect\normalsize Conclusion}
       We propose the implementation of all-optical reversible universal logic gates (which are relevant to build a reversible processor), by application of adiabatic population transfer, based on STIRAP-like processes. In particular, we demonstrate a three-bit and a four-bit Toffoli gate. The three-bit Toffoli gate is implemented in a three-level $\Lambda$-type level scheme, driven by a pump and a Stokes laser pulse. The three input bits of the gate are defined by the two laser pulses and the state of the system at the beginning of the interaction. The output state is defined by the state of the system at the end of the interaction. The same pulse sequence, with laser frequencies tuned to sufficiently large single-photon detunings of the relevant transitions serves to completely drive atomic population from an initial state to final state and back again without losses (i.e. in a reversible way). The truth table of the interaction (or the optical bits, as defined by the laser pulses and the atom, respectively) resembles a three-bit Toffoli gate.
The four-bit Toffoli gate is implemented in a five-level system (e.g. an extended $\Lambda$-type system or a M-type scheme), driven by three laser pulses on four transitions.The four input bits of the gate are defined by the three laser pulses and the state of the system at the beginning of the interaction. The output state is defined by the state of the system at the end of the interaction. When pairs of laser frequencies are tuned to two-photon resonance in the five-level systems, the schemes essentially reduce to an effective $\Lambda$-system, where the ground states are superposition states. We derive the dressed states and dressed energies of the system, as well as conditions for adiabatic evolution of the population dynamics. We show, that adiabatic passage permits reversible transfer of atomic population from an initial to a target state, and back again. The truth table of the interaction (or the optical bits, as defined by the laser pulses and the atom, respectively) resembles a four-bit Toffoli gate.

Finally, we derive restrictions for implementation of adiabatically-driven Toffoli gates in optically dense media. We find, that the adiabaticity is preserved, if the length of the medium does not exceed the linear absorption length.

\section*{Acknowledgments}
The research leading to these results has received funding from Deutsche Forschungsgemeinschaft,
the Volkswagen foundation, and the People Programme (Marie Curie Actions) of the European Union's
Seventh Framework Programme FP7/2007-2013/ under REA Grant Agreement No. 287252 and 295025.We acknowledge additional support from the IRMAS International Associated Laboratory (CNRS-France SCS-Armenia).

\bigskip

\section*{Appendix A}

We derive now the eigenvalues and conditions for adiabatic population dynamics in the five-level system,as studied in section III.
From the equation for the eigenvalues of the Hamiltonian $det(H -\Lambda I)=0$ and assuming
$\delta_2=\delta_4=0$ and $\delta_1=\delta_3=\Delta$ we get
\begin{eqnarray}
\Lambda^2(\Lambda-\Delta)[\Lambda (\Lambda-\Delta)+\Omega^2_s]+V^4\Lambda =0
\end{eqnarray}
with $\Omega^2_s=\Omega^2_1+\Omega^2_2+\Omega^2_3+\Omega^2_4$ and $V^4=\Omega^2_2\Omega^2_4+\Omega^2_1\Omega^2_3+\Omega^2_1\Omega^2_4$. Using the notation $x=\Lambda(\Lambda-\Delta)$, the above equation becomes $\Lambda[x^2-\Omega^2_s x + V^4]=0$ and the eigenvalues are :
\begin{eqnarray}
\Lambda_0=0,\\
\Lambda_{3,1}=\frac{1}{2}[\Delta\pm(\Delta^2+4x_1)^{1/2}],\\
\Lambda_{4,2}=\frac{1}{2}[\Delta\pm(\Delta^2+4x_2)^{1/2}]
\end{eqnarray}
where $x_{2,1}=(1/2)[\Omega^2_1\pm(\Omega^4_s-4V^4)^{1/2}]$. We note that the condition $\Omega^4_s\geq 4V^4$ is always met.

Adiabatic evolution for a single atom requires $|\Lambda_i-\Lambda_j|T\gg1$ for any $i\neq j$ with the interaction time $T$. This leads to the conditions for the pulse parameters
\begin{eqnarray}
\frac{(x_2-x_1)T}{(\Delta^2+4x^2_2)^{1/2}}\gg1,\\
(\Delta^2+4x_1)^{1/2}T\gg 1,\\
\frac{x_{1,2}T}{(\Delta^2+4x_{1,2})^{1/2}}\gg 1
\end{eqnarray}
Note that the last condition can be fulfilled only for $V^4\neq0$, i.e. for temporal overlap of the pulses.

With $\Omega^2_1=\Omega^2_4$ the expressions for $x_{1,2}$ are simplified and yield $x_1=\Omega^2_1, x_2=\Omega^2_s,$ and $\Omega^4_s-4V^4=\Omega^4.$

In this case the adiabaticity conditions are
\begin{eqnarray}
\frac{(\Omega^2_2+\Omega^2_3)T}{(\Delta^2+4\Omega^2_s)^{1/2}}\gg1,\\
(\Delta^2+4\Omega^2_1)^{1/2}\gg 1,\\
\frac{\Omega^2_1 T}{(\Delta^2+4\Omega^2_1)^{1/2}}\gg 1,\\
\frac{\Omega^2_s T}{(\Delta^2+4\Omega^2_s)^{1/2}}\gg 1
\end{eqnarray}

\section*{Appendix B}
We briefly discuss now the basic theoretical treatment of pulse propagation effects and their effect upon adiabaticity in our five-level systems, as discussed in section III.
By combining the Schr\"{o}dinger equation with the truncated Maxwell equation we obtain for a medium of five-level M-type atoms a self-consistent system of equations. These describe variations of pulse frequencies (i.e. single-photon detunings) and intensities (i.e. Rabi frequencies), when the pulses propagate through the medium:
\begin{eqnarray}\label{eq9}
\frac{\partial\Omega_1}{\partial z}=q_1\frac{\partial}{\partial \tau}|b_1|^2+q_4\frac{\partial}{\partial \tau}|b_5|^2\nonumber \\
\frac{\partial\Omega_2}{\partial z}=-q_2\frac{\partial}{\partial \varepsilon}(|b_1|^2+|b_2|^2), \nonumber \\
\frac{\partial\Omega_3}{\partial z}=-q_3\frac{\partial}{\partial \varepsilon}(|b_4|^2+|b_5|^2), \nonumber \\
\frac{\partial\Delta_1}{\partial z}=q_1\frac{\partial}{\partial \tau} \frac{Re(b^*_1b_2)}{\Omega_1}+q_4\frac{\partial}{\partial \tau} \frac{Re(b^*_4b_5)}{\Omega_1}, \nonumber \\
\frac{\partial\Delta_{2,3}}{\partial z}=-q_{2,3}\frac{\partial}{\partial \tau} \frac{Re(b^*_{2,3}b_{3,4})}{\Omega_1}
\end{eqnarray}
Here  $q_i=2\pi\omega_i|d_{i,i+1}|^2N/(\hbar c)$, $b_i(z,\tau)$ are amplitudes of atomic populations in dressed states. z,$\tau$ are running coordinates $z=x$, $\tau=t-x/c$. In the case of an extended $\Lambda$-system (Fig.1b) the equations essentially remain the same, but we must exchange signs in $\Omega^2_2$ and $\Omega^2_3$.
We note, that self-phase modulation may lead to variations of frequencies and, hence detunings from corresponding resonances (i.e. parametric broadening of the pulse spectrum \cite{35 Y.R.Shen}). The modification of the pulse shapes is caused both by the nonlinear group velocity (which can result in formation of shock wavefronts \cite{36 D.Grischkowsky}) and by energy transfer between the pulses (which can lead to full depletion of one of the pulses \cite{33 Chaltykyan}). The strength of all these processes depends upon the optical length. Hence, if the optical length of the medium is sufficiently short, the variations of detunings and intensities can be negligibly small to enable adiabatic evolution.
A detailed discussion of pulse propagation dynamics in our five-level system would extend the present paper too much. Here, our specific aim is to define a restriction for the medium length, which still permits adiabatic evolution, i.e. proper formation of adiabatic states $|\Lambda_1\rangle$ and $|\Lambda_2\rangle$ in all atoms of the medium.
Since it is the time derivatives that enter the right-hand sides of equations (13), we can use for atomic populations the expressions following from those for the states $|\Lambda_1\rangle$ and $|\Lambda_2\rangle$. This means taking into account the first nonadiabatic corrections.
If the changes in the Rabi frequencies and detunings (as given by the right hand side of equations (13)) are sufficiently small, adiabaticity will be maintained over the full medium length.
Estimating the right-hand sides of equations  (26), we arrive at the restricting conditions (4) and (13).

\end{document}